\documentclass{PoS}

\title{The Dirac operator spectrum: a perturbative approach}

\ShortTitle{The Dirac operator spectrum: a perturbative approach}

\author{M.~Brambilla\\
Universit\`a di Parma \& INFN, Viale Usberti 7/A, I-43100 Parma, Italy\\
E-mail: \email{michele.brambilla@fis.unipr.it}}

\author{\speaker{F.~Di~Renzo}\\
Universit\`a di Parma \& INFN, Viale Usberti 7/A, I-43100 Parma, Italy\\
E-mail: \email{francesco.direnzo@fis.unipr.it}}

\abstract{By computing the Dirac operator spectrum by means of Numerical Stochastic Perturbation 
Theory, 
we aim at throwing some light on the widely accepted picture for the mechanism which is behind 
the Bank-Casher relation. The latter relates the chiral condensate to an accumulation of 
eigenvalues in the low end of the spectrum. This can be in turn ascribed to the usual mechanism 
of repulsion among eigenvalues which is typical of quantum interactions.
First results appear to confirm that NSPT can indeed enable us to inspect a huge reshuffling of 
eigenvalues due to quantum repulsion.}

\FullConference{The XXVII International Symposium on Lattice Field Theory - LAT2009\\
		 July 26-31 2009\\
		 Peking University, Beijing, China}

\begin{document}

\section{Introduction}

Chiral symmetry breaking is one the key feature of QCD. There is a very intuitive way of stating the 
physics of this phenomenon: a small quark mass leads to a macroscopic realignment of the QCD 
vacuum (this is a strict quotation from~\cite{JVTilo}). 
Since the QCD partition function reads

\begin{equation}\label{eqn:Z}
  Z = \langle \prod_f \det ({\cal D} + m_f) \rangle = \langle \prod_f \prod_n (i \lambda_n + m_f) \rangle
\end{equation}

\noindent
in order for this to be possible there must be an accumulation of 
Dirac operator eigenvalues near zero (otherwise the effect of a small quark mass would be 
overwhelmed 
by much larger eigenvalues). This message is actually encoded in the Banks Casher 
relation~\cite{BaCa}

\begin{equation}\label{eqn:BC}
  \langle \bar{\psi} \psi \rangle = \frac{\pi \rho(0)}{V}
\end{equation}

\noindent
relating the chiral condensate (the order parameter of the transition associated to spontaneous 
symmetry breaking) to the density of eigenvalues of the Dirac operator spectrum

\begin{equation}\label{eqn:rho}
  \rho(\lambda) = \langle \sum_n \delta (\lambda - \lambda_n) \rangle.
\end{equation}

Altought not a natural observable in Field Theory, the Dirac operator spectrum has in force 
of~(\ref{eqn:BC}) become a natural probe for the chiral transition. Recent work~\cite{LeoLusch} 
has investigated the field theoretic status of spectral observables, in particular with respect 
to their renormalization properties. From a numerically point of view, it should be pointed out 
that Lattice QCD can quite naturally compute~(\ref{eqn:rho}), once a lattice 
regularization of the Dirac operator is given.

\section{The Dirac spectrum and Perturbation Theory}

Since the free Dirac operator has a vanishing eigenvalues density near zero, one is lead to the 
conclusion that the small eigenvalues are due to gauge interactions. There is actually a natural 
candidate: any quantum interaction produces a repulsion among the eigenvalues. With this respect 
Perturbation Theory is in a tantalizing situation:
\begin{itemize}
\item
on one side, it sits (deep) in the chirally restored regime, while one looks for an effect 
which lives at its border;
\item
on the other side, it gives a unique opportunity to follow the fate of eigenvalues in their 
mutual repulsion.
\end{itemize}

We want to emphasize that our work is still at a very preliminary stage. In particular, we 
do not want to address here the subtleties which arise in properly defining a perturbative 
expansion of~(\ref{eqn:rho}). We will discuss a \emph{quick and dirty} procedure in which 
we first compute the perturbative corrections to the free spectrum eigenvalues
\[
\lambda_n = \lambda_n^{(0)} + \beta^{-1/2} \lambda_n^{(1)} + \beta^{-1}\lambda_n^{(2)} + \ldots
\]
and then resum the expansion at given values of the coupling $\beta$. Given these summations, 
we can proceed to compute a density of eigenvalues much the same as in non-perturbative 
computations of the spectrum.

\section{The Dirac Spectrum in NSPT}

Numerical Stochastic Perturbation Theory~\cite{NSPT} relies on an expansion of the solution of 
Langevin equation. In the case of LGT

\begin{equation}\label{eqn:NSPT}
  U_{x\mu}(\tau;\eta) \rightarrow 1+ \sum_{k=1} \beta^{-k/2} U^{(k)}_{x\mu}(\tau;\eta).
\end{equation}

\noindent
$\tau$ is the stochastic time of Langevin evolution, with 
gaussian noise $\eta$. For asymptotic values of the stochastic time, 
$\eta$-averages $\langle \ldots \rangle_{\eta}$ of observables converge order by order to quantum 
field theory averages $\langle \ldots \rangle_{QFT}$.\\

Plugging~(\ref{eqn:NSPT}) into the Dirac operator turns the computation of~(\ref{eqn:rho}) into the 
typical eigenvalue/eigenvector problem in PT

\begin{equation}
  M = M_0 + N = M_0 + \sum_i g^i N_i \;\;\;\;\;\;\;\;\; M \, |\alpha\rangle = \epsilon \, |\alpha\rangle
\end{equation}
which has to be solved by
\begin{equation}
\epsilon = \epsilon_0 + g \, \epsilon_1 + g^2 \, \epsilon_2 + \ldots \;\;\;
|\alpha\rangle = |\alpha_0\rangle + g \, |\alpha_1\rangle + g^2 \, |\alpha_2\rangle + \ldots
\end{equation}

\noindent
Due to the (huge) degeneracy of the free field solution, for every eigenvalue we need to explicitly 
separate 
components inside and outside the starting (degenerate) eigenspace, \emph{i.e.}

\begin{equation}
|\alpha\rangle = |\alpha_0\rangle + P'_{in} |\alpha\rangle + P_{out} |\alpha\rangle.
\end{equation}

\noindent
In the previous formula $|\alpha_0\rangle$ is the direction in the free (degenerate) eigenspace 
singled out as the zeroth order of the solution; $P'_{in}$ is the projector onto the 
component of the free eigenspace which is orthogonal to $|\alpha_0\rangle$;
$P_{out}$ projects instead outside the free eigenspace. 
We finally get the iterative solution 
\begin{eqnarray}\label{eqn:solution}
\epsilon_n &=& \sum_{k=0}^{n} \, \langle \alpha_0 |N_{n-k}| \alpha_k\rangle \\ \nonumber
P_{out} | \alpha \rangle &=& (\epsilon - M_0 - P_{out}N )^{-1} \left( \, P_{out}N |\alpha_0 \rangle + P_{out}N P'_{in}
|\alpha \rangle \, \right)\\ \nonumber
P'_{in} | \alpha \rangle &=& \; (\epsilon - \epsilon_0 - P'_{in}N )^{-1} \;\; \left( \, P'_{in}N |\alpha_0 \rangle + 
P'_{in}N P_{out}|\alpha \rangle \, \right).\\ \nonumber
\end{eqnarray}
This is the (closed) solution only provided degeneracy is lifted at first order. Should this not 
be the case, the formalism should be generalized by introducing a new projector for each level 
of degeneracy still present (the solution is nevertheless closed also in such a situation, 
which actually occurs in our computations). \\
 
In standard non-perturbative LGT computations of the Dirac operator spectrum one gets distributions 
of eigenvalues by generating configurations and computing the spectrum on each of them. 
The density of eigenvalues is then simply obtained by plain histograms of the results. We stress 
once again that at this stage of our work we will adhere to the \emph{naive} recipe of first 
computing the eigenvlaues in PT, then summing the expansions at given values of the coupling 
and finally constructing histograms much the same way as in the non-perturbative case.

\section{Results}

In figure~(1) we plot examples of our results: we collect all the measurements for first (trivial) 
and second (one loop) order corrections to free field results for the second lowest lying eigenspace 
on a $6^4$ lattice. We stress that this eigenspace is degenerate (the dimension of this eigenspace is 
$144$), but on top of this degeneracy the histograms entail the multeplicity which comes from the 
number of measurements. 
\begin{figure}[!htb]
\begin{center}
\includegraphics[scale=0.62,clip=true]
{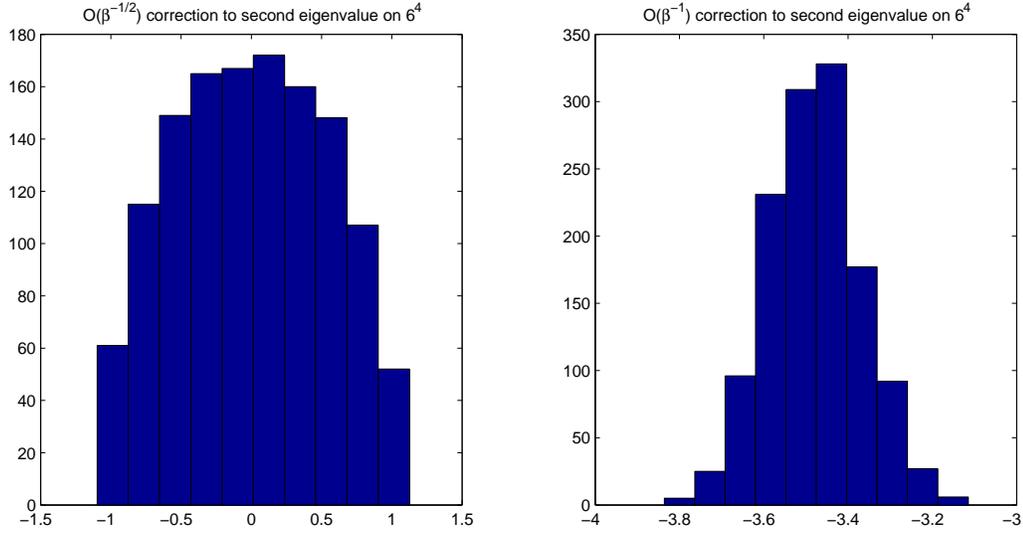}
\end{center}
  \vspace{-5mm}
  \caption{First (trivial) and second (one loop) corrections to the second (lowest lying) free field eigenvalue on a $6^4$ lattice (overall distributions of the measures).}
  \label{fig:}
\end{figure}

\begin{figure}[!htb]
\begin{center}
\includegraphics[scale=0.62,clip=true]
{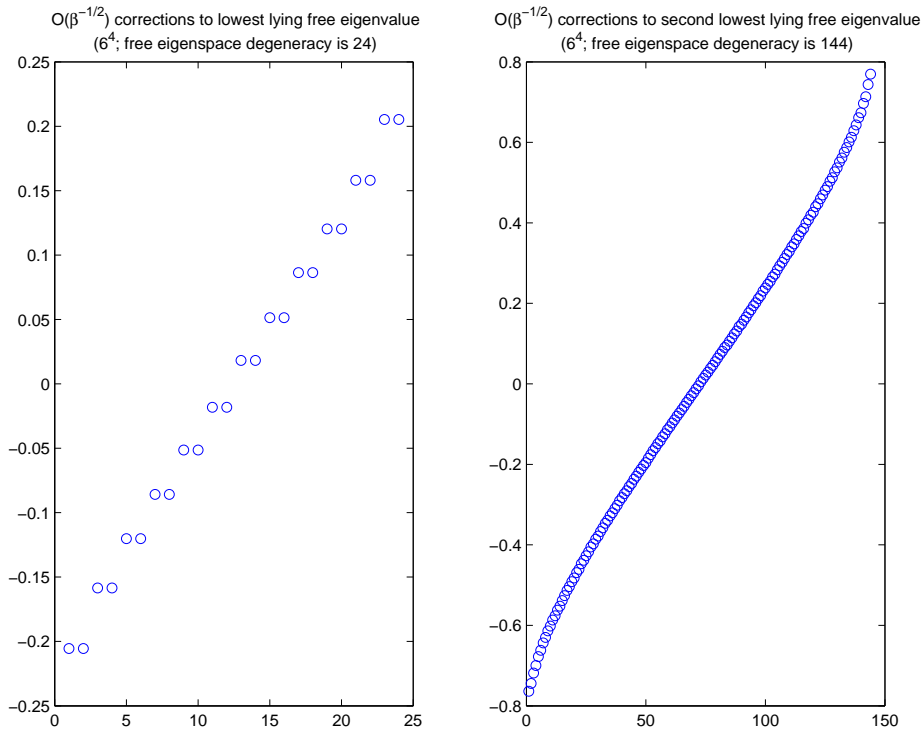}
\end{center}
  \vspace{-5mm}
  \caption{First (trivial) corrections to the first (lowest lying) and second free field eigenvalue on a $6^4$ lattice (averages over Langevin histories). Free eigenspace degeneracies are 24 (left) and 144 (right).}
  \label{fig:}
\end{figure}

Figure~(2) displays data once the average over all the measurements has been taken. In this case we plot 
first order corrections (as one expects, they average to zero) for lowest lying and second lowest lying 
eigenvalues. There are issues which are worth stressing. First of all, one can ispect degeneracies which 
are not lifted. Second, the distributions of corrections in the two eigenspaces differ quite a lot.\\

Figure~(3) displays another interesting feature. In this case we plot a third order correction, which 
enlights how higher orders display long tails. One probably needs to 
carefully assess when the free field degeneracy is actually lifted, as it is clear from the impact of 
denominators in~(\ref{eqn:solution}).\\

\begin{figure}[!htb]
\begin{center}
\includegraphics[scale=0.62,clip=true]
{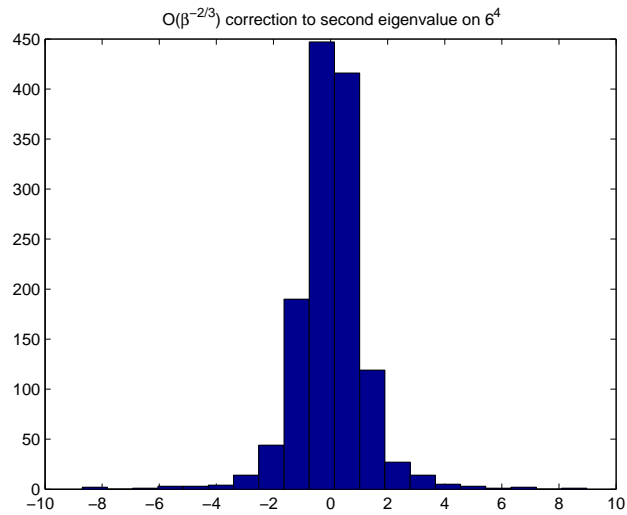}
\end{center}
  \vspace{-5mm}
  \caption{Third correction to the second free field eigenvalue on a $6^4$ lattice: while it is centered 
in zero (as expected), it displays long tails.}
  \label{fig:}
\end{figure}

With this respect we point out that one can always check the accuracy of the computation by 
considering quantities like 

\[
\;\;\;\;\;\;\;\; \langle \mbox{Tr} (D^\dagger D)^k \rangle = \ldots \;\;\;\;\;\;\;\; 
\langle \mbox{Tr} (D^\dagger D)^{-k} \rangle = \ldots
\]

They can be both computed directly and reconstructed from the eigenvalues distribution, eventually 
validating the latter. \\
 
We can now go back to our \emph{quick and dirty} procedure to inspect the impact of the perturbative 
corrections. Basically, we can 
sum the contributions at any given value of the coupling and try to follow the resulting 
distribution of eigenvalues as the gauge intercation comes into play. We plot in figure~(4) 
what we get at one loop.

\begin{figure}[!htb]
\begin{center}
\includegraphics[width=\textwidth,height=12cm,clip=true]
{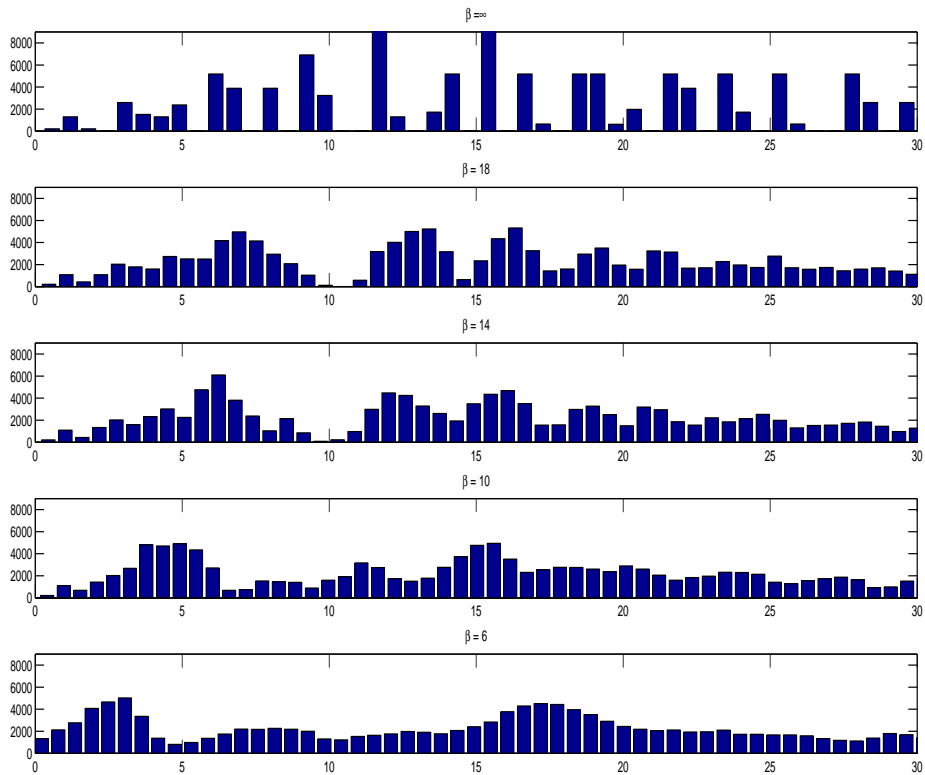}
\end{center}
  \vspace{-5mm}
  \caption{The evoultion of the eigenvalues density: from free field limit ($\beta=\infty$)
to the intercating case (at different values of the coupling $\beta$).}
  \label{fig:}
\end{figure}

Figure~(4) is something like a sequence of pictures taken while the interaction is switched on. One 
starts at zero coupling, where the key feature of the free field is on display: bins are centered 
where free field eigenvalues sit, and bins heigth simply entails the degeneracy of the various 
eigenspaces. Notice anyway that at this resolution some bins actually results from 
the contribution of two free field eigenvalues sitting very close to each other. 
While the interaction is switched on (\emph{i.e.} the value of the inverse coupling $\beta$ decreases) 
the bins spread and overlap and eventually a non-zero density near zero is generated. 
A natural question arises: where do eigenvalues moving to zero come from? 
One should remember the point we made on repulsion among eigenvalues. Figure~(5) displays an example 
of how this takes place: we plot 
the contribution to $\rho$ coming from two eigenvalues starting very close to each other in free field.\\

It is worth better assessing the impact of the repulsion among the couple of free eigenvalues we have just 
looked at. It 
actually turns out that they give a substantial contribution to the rearrangement of the 
eigenvalues density: one can recognize their splitting on the right of figure~(6).

\begin{figure}[!htb]
\begin{center}
\includegraphics[width=\textwidth,height=11.5cm,clip=true]
{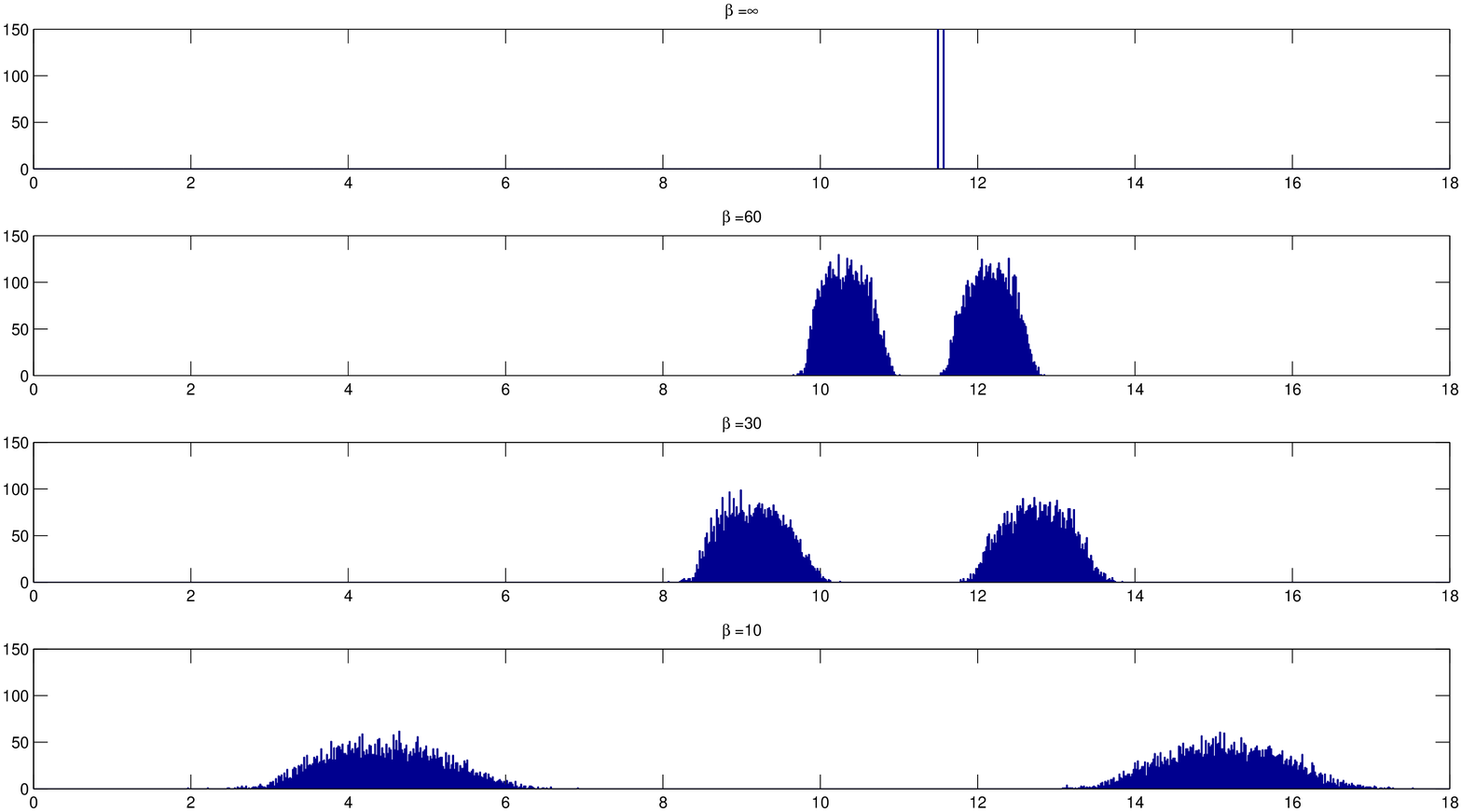}
\end{center}
  \vspace{-5mm}
  \caption{Following the repulsion of two eigenvalues on $6^4$. They start very close 
in free field limit and then strongly repel each other.}
  \label{fig:}
\end{figure}

\begin{figure}[!hbt]
\begin{center}
\includegraphics[scale=0.62,clip=true]
{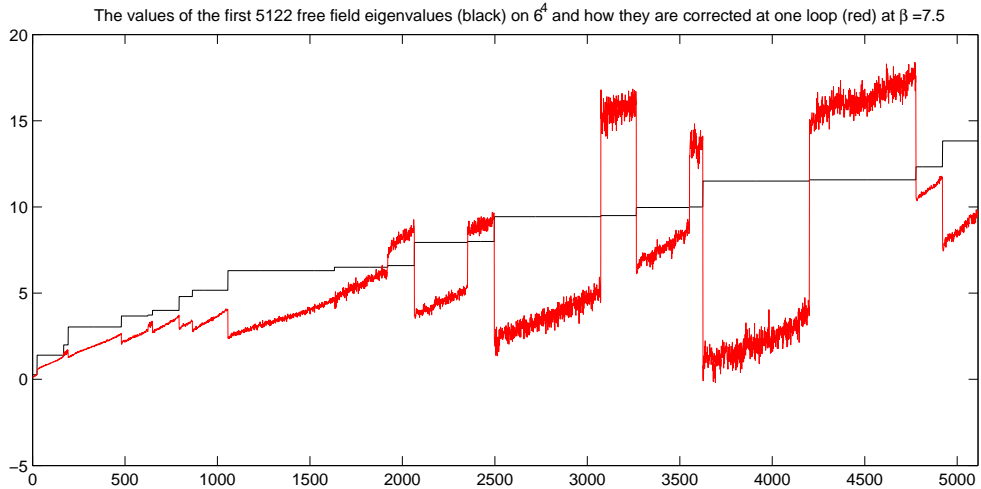}
\end{center}
  \vspace{-5mm}
  \caption{The first 5122 free eigenvalues on a $6^4$ lattice (black line): the lenghts of each 
segment is the degeneracy in free field. Red curve displays how they move at one loop at 
$\beta=7.5$.}
  \label{fig:}
\end{figure}

Black line in figure~(6) is nothing but another way of plotting the first row of figure~(4): we plot 
the first 5122 free eigenvalues and the lenght of each plateaux is just the degeneracy of each free 
field eigenspace. The superimposed red line shows the summation (at first loop) of the perturbative 
series for these eigenvalues at $\beta=7.5$.\\

Some caveats are of course in order:
\begin{itemize}
\item
Is this a finite-volume effect? At the moment we have actually got 
the same qualitative picture at any (still moderate) size we studied.
\item
Is this a finite $a$ effect? Testing this is more difficult.
\item
One should carefully take care of the order of limits which is in place 
in the Banks Casher relation.
\end{itemize}

A few following steps are on their way: we will repeat the computation in the background of different 
$Z(3)$ vacua and to try to reconstruct the Polyakov loop from the spectral decomposition 
of the Dirac operator (this is in the spirit of recent works by Gattringer~\cite{CGatt}). 

\section{Conclusions}

Even though at a very preliminary stage, we showed some results of a perturbative computation 
of the Dirac operator spectrum by means of NSPT.
Our results quantitatively support the picture of the repulsion among eigenvalues being responsible 
for the rearrangement of eigenvalues, ultimately giving rise to Banks Casher.\\

Some developments of this work are expected to follow these preliminary results.

\begin{itemize}
\item
We have to carefully assess the huge tails of higher order distributions. One hint is 
that this asks for some regulator in highly degenerate free field eigenspaces.
\item
We will move to computations in the background of different $Z(3)$ vacua.
\item 
Having at hand full spectra could in principle enable the computation of a variety of quantities.
\end{itemize}

\end{document}